# Thermodynamics of Non-equilibrium Diffuse-Interfaces in Mesoscale Phase Transformations

Yue Li, Lei Wang∗, Junjie Li, Jincheng Wang, Zhijun Wang∗

State Key Laboratory of Solidification Processing, Northwestern Polytechnical University, Xi'an 710072, China

*Abstract*.

We present a new phase-field formulation for the non-equilibrium interface kinetics. The diffuse interface is considered an integral of numerous representative volume elements (RVEs), in which there is a two-phase mixture with two conserved and two non-conserved concentration fields. This particular way of separating concentration fields leads to two distinct dissipative processes besides the phase boundary migration, i.e., trans-sharp-interface diffusion between different phases within a single RVE and trans-diffuse-interface diffusion between different RVEs within a single phase. A "two-part" mechanism is proposed to balance the driving forces and energy dissipation for diffusionless and diffusional processes in RVEs. Moreover, the classical interface friction and solute drag can be recovered by integrating the diffuse interface. Compared with the sharp interface theory, the present diffuse interface can reproduce the atomic simulated "partial-drag" self-consistently in thermodynamics.

---

∗ Corresponding author. lei.wang@nwpu.edu.cn (L. Wang)
∗ Corresponding author. zhjwang@nwpu.edu.cn (Z. Wang)

*Motivation*.—Many inhomogeneous systems at the mesoscale consist of domains of well-defined phases separated by distinct interfaces [1]. Due to the coexistence of individual and cooperative atomic motions [2-6], the dissipative processes within the interfaces are usually more complex than those in the bulks. Therefore, to quantitatively predict phase transformations, it is vital to uncover how various dissipative processes partition the thermodynamic forces [6] and develop suitable models for describing interface kinetics from near to far from equilibrium.

As the most fundamental approach, classical sharp interface models can be directly applied to transformation kinetics with simple morphology [7] and serve as the standard or rule for other simulation techniques, e.g., phase-field modeling, cellular automation, and level-set methods [8]. However, sharp interface models fail in the kinetics of interfaces far from equilibrium. A simple but illustrative example is the unusual "partial drag" [9] in the rapid solidification of highly undercooled melts. Specifically, although sharp interface models have been extended to non-equilibrium cases [10-14] based on the classical irreversible thermodynamics [15-17], the predicted driving force and energy dissipation significantly deviate from those of experiments [18] and atomic simulations [19-22]. Therefore, a thermodynamically consistent description of far-from-equilibrium interfaces is still lacking.

The failure of sharp interface models is likely attributed to the diffuse nature of many interfaces. The phase-field (PF) theory has become a mainstream means of modeling diffuse interfaces by introducing a series of fields representing the material's properties with the interface [23, 24]. As the continuum limit of microscopic models [24], the classical mesoscale PF models should inherently reflect the averaged properties of diffuse interfaces. However, mainly focusing on the sharp interface limit [25-27], they are more like an effective sharp interface model for capturing complex pattern formations. By contrast, there is no clear understanding of the dissipative processes and their thermodynamics inside the diffuse interface. Therefore, this letter aims to develop a PF model self-consistently in thermodynamics, with transparent dissipative processes inside the diffuse interface to represent the far-from-equilibrium kinetic.

*Model*.— This model starts from a classical free energy formulation of a two-phase binary system, i.e., $F = \int \left\{ \sum_{i=\alpha,\gamma} h_i(\phi)(f_i/v_m) + wg(\phi) + \frac{\kappa^2}{2}(\nabla \phi)^2 \right\} d\Omega$, in which $h_i(\phi) = \phi_i^2(3-2\phi_i)$ is used to interpolate the free energy density $f_i/v_m$, $w$ and $\kappa^2$ are coefficients for the double-well potential $g(\phi) = \phi^2(1-\phi)^2$ and gradient energy, respectively. The overall concentration is $c = h_\alpha c_\alpha + h_\gamma c_\gamma$.

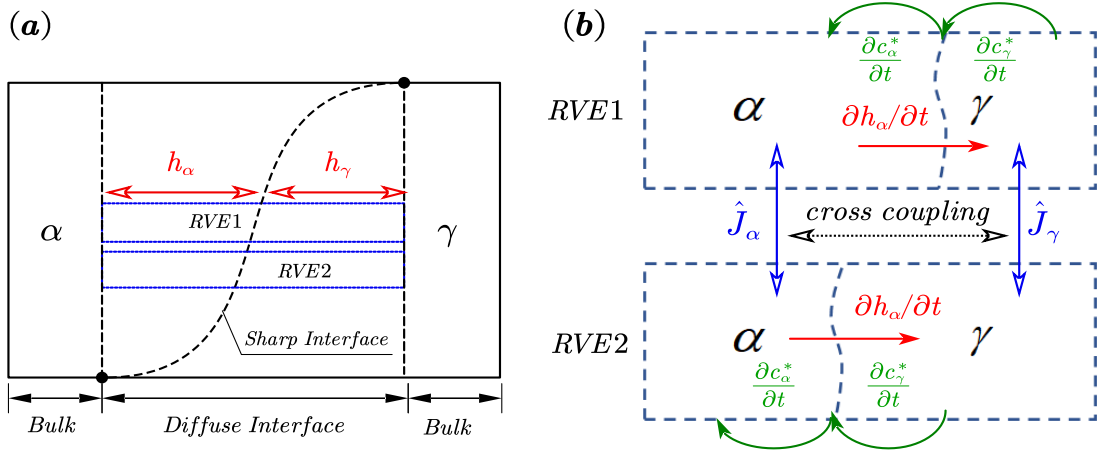

FIG. 1. (a) Diffuse interface and the internal representative volume elements (RVEs); (b) Three dissipative processes between and within RVEs, where the blue arrows means the long-range diffusion in the same phase while the green arrows are the short-range exchange between different phases.

As shown in Figure. 1(a), the diffuse interface is considered an integral of numerous representative volume elements (RVEs), in which "RVE1" and "RVE2" are adjacent. The dashed curve denotes the average phase fractions rather than orderly arrangement of two phases. For a single RVE in Figure. 1(b), the domains of $\alpha$ and $\gamma$ are separated by a distinct phase boundary as a "sharp interface", whose kinetics is controlled by phase boundary migration (red arrows) and short-range atomic exchange (green arrows), referred to as "trans-sharp-interface diffusion" in this work. Furthermore, between

3 / 12

adjacent RVEs, there is a long-range diffusion in the same phase (blue arrows), which is defined as "trans-diffuse-interface diffusion" currently.

Due to the atomic exchange, it is natural to see that the separated concentration fields $c_{i=\alpha,\gamma}$ should be non-conserved. Thus, we here divide the instantaneous $c_i$ into two fields: a conserved $\hat{c}_i$ and a non-conserved $c_i^*$, leading to a diffusion equation with additional sources or sinks, i.e.,

$$\frac{\partial c_i}{\partial t} = -v_m \nabla \hat{J}_i + \frac{\partial c_i^*}{\partial t}, \qquad (1)$$

where $\hat{J}_i$ and $\partial c_i^*/\partial t$ ($i = \alpha, \gamma$) are the fluxes of trans-diffuse-interface diffusion and trans-sharp-interface diffusion in Figure. 1(b), respectively. Note that it is the trans-sharp-interface diffusion, together with phase boundary migration, that determine the "sharp-interface" phase transformation. Therefore, there should be $\hat{J}_\alpha = \hat{J}_\gamma = J$ (or $\hat{c}_\alpha = \hat{c}_\gamma = c$), making the mass conservation $c = h_\alpha c_\alpha + h_\gamma c_\gamma$ yield

$$h_\alpha \frac{\partial c_\alpha^*}{\partial t} + h_\gamma \frac{\partial c_\gamma^*}{\partial t} = -\frac{\partial h_\alpha}{\partial \phi} \frac{\partial \phi}{\partial t}(c_\alpha - c_\gamma), \qquad (2)$$

which is also similar to the classical sharp interface models [9-14]. Addition, it should be note that a similar understanding of internal dissipative processes has been previously used in phase-field models [28, 29], but the present Eqs. (1) and (2) are treating their kinetic coupling in an exactly different way.

Now, let us consider a phase transformation in a closed system, the change rate of total free energy $\dot{F}$ is given by [30]

$$\dot{F} = \int_\Omega \left\{ \frac{\partial \phi}{\partial t} \frac{\delta F}{\delta \phi} + J \cdot \sum_{i=\alpha,\gamma} v_m \nabla \frac{\delta F}{\delta c_i} + \sum_{i=\alpha,\gamma} \frac{\partial c_i^*}{\partial t} \frac{\delta F}{\delta c_i} \right\} d\Omega \\ + \lambda \int_\Omega \left\{ h_\alpha \frac{\partial c_\alpha^*}{\partial t} + h_\gamma \frac{\partial c_\gamma^*}{\partial t} + \frac{\partial h_\alpha}{\partial \phi} \frac{\partial \phi}{\partial t}(c_\alpha - c_\gamma) \right\} d\Omega, \qquad (3)$$

where $\lambda$ is the Lagrange multiplier for the constraint of Eq. (2). Based on the extremum principle of linear thermodynamics [17], the total free energy dissipation $Q$ is then expressed as a quadratic form of four fluxes in Eq. (3)

$$Q = \int_\Omega \left\{ \frac{1}{M_\phi}\left(\frac{\partial \phi}{\partial t}\right)^2 + \frac{\hat{J}^2}{M_c} + \sum_{i=\alpha,\gamma} \frac{1}{M_i^*}\left(\frac{\partial c_i^*}{\partial t}\right)^2 \right\} d\Omega, \qquad (4)$$

where $M_\phi$ is the phase-field mobilities, $M_c$ and $M_i^*$ are the atomic mobilities for $J$ and $\partial c_i^*/\partial t$, respectively. Combining Eq. (2) with $\delta(\dot{F}+Q/2)=0$ for four fluxes [17] can eliminate the $\lambda$ and solve the evolving equations as ($i=\alpha,\gamma$, $j \neq i$) [30]

$$\frac{1}{M_\phi^{eff}}\frac{\partial \phi}{\partial t} = \kappa^2 \nabla^2 \phi - w\frac{\partial g}{\partial \phi} - \frac{1}{v_m}\frac{\partial h_\alpha}{\partial \phi}\left[ f_\alpha - f_\gamma - (h_\alpha \tilde{\mu}_\alpha + h_\gamma \tilde{\mu}_\gamma)(c_\alpha - c_\gamma) \right], \qquad (5)$$

$$\frac{\partial c_i^*}{\partial t} = \frac{M^* h_\alpha h_\gamma}{v_m h_i}(\tilde{\mu}_j - \tilde{\mu}_i) + \frac{\partial h_\alpha}{\partial \phi}\frac{\partial \phi}{\partial t}(c_\gamma - c_\alpha), \qquad (6)$$

$$J = \hat{J}_i = -M_c \nabla(h_\alpha \tilde{\mu}_\alpha + h_\gamma \tilde{\mu}_\gamma), \qquad (7)$$

in which $\tilde{\mu}_i$ are the diffusion potentials, $M_\phi^{eff} = \left[ \frac{1}{M_\phi} + \left(\frac{\partial h_\alpha}{\partial \phi}\right)^2 \frac{(c_\alpha - c_\gamma)^2}{M^*} \right]^{-1}$ is an effective phase-field mobility, $M_c = \left(\sum_i h_i D_i\right)\left[\sum_i h_i (\partial \tilde{\mu}_i/\partial c_i)\right]^{-1}$ and $M^* = M_i^*/h_i = (5.88 v_m^2 M_c)/(2a\delta)$ are the atomic mobilities for $\hat{J}_i$ and $\partial c_i^*/\partial t$, respectively [30], where $a$ is the lattice distance and $2\delta$ is the interface width. Additionally, one can see that Eqs. (5)-(7) follow the Onsager's reciprocal relationship [15, 30]. Different from existing models [28, 29], the present Eq. (7) indicates that the trans-diffuse-interface diffusion of $\alpha$ is also driven by the chemical gradient of another because $c_\alpha$ is being changed by the trans-sharp-interface diffusion (vice versa). Thus, $\hat{J}_\alpha$ and $\hat{J}_\gamma$ are dependent in Figure. 1. Substituting Eqs. (6) and (7) into Eq. (1) will offer the temporal evolutions of $c_\alpha$, $c_\gamma$, and $c$ [30].

*Thermodynamics of sharp-interface*.—In a single RVE, there is a "sharp-interface" type phase transformation governed by the response functions of interface velocity [Eq. (5)] and concentrations [Eq. (6)]. According to Eq. (2), the fluxes of trans-sharp-interface diffusion are given by ($i=\alpha,\gamma$)

$$\frac{\partial c_i^*}{\partial t} = \frac{1}{h_i}\frac{\partial h_i}{\partial \phi}\frac{\partial \phi}{\partial t}\left(c_{trans}^{SI} - c_i\right), \tag{8}$$

where $c_{trans}^{SI} = c_i + h_i\left(\partial c_i^*/\partial t\right)/\left(\partial h_i/\partial t\right)$ is the material composition crossing the sharp interface, instead of the real solid composition $c_i$ [12]. Substituting Eq. (8) into the non-conserved terms of Eq. (3), the net driving force $\Delta f_{tot}^{SI}$ for a RVE is

$$\Delta f_{tot}^{SI} = f_\alpha - f_\gamma + \tilde{\mu}_\alpha\left(c_{trans}^{SI} - c_\alpha\right) + \tilde{\mu}_\gamma\left(c_\gamma - c_{trans}^{SI}\right), \tag{9}$$

which is the same as the sharp-interface models [11-15]. It means that $\Delta f_{tot}^{SI}$ is consumed to drive a diffusional transformation from $\gamma$ to $\alpha$ and two fluxes being transferred, for which "$f_\alpha - f_\gamma$", "$\tilde{\mu}_\gamma$", and "$\tilde{\mu}_\alpha$" are three conjugated forces cross-coupled with three fluxes [15]. However, in order to distinctly understand different dissipative processes and their thermodynamics, it is necessary to find independent fluxes and forces.

Considering a steady state in RVEs, the independent driving force for phase boundary migration $\Delta f_m$ according to Eq. (5) yields as

$$\Delta f_m = \sum_{i=\alpha,\gamma} h_i\left[f_\alpha - f_\gamma - \tilde{\mu}_i\left(c_\alpha - c_\gamma\right)\right]. \tag{10}$$

It suggests that there are two types of diffusionless transformation for phase boundary migration, and only the part $h_\alpha$ migrates as Hillert et al.'s viewpoint [6]. Thus, a "two-part" concept will be used in latter discussions, i.e., Part I ($h_\alpha$) and Part II ($h_\gamma$). Substituting $\partial h_\alpha/\partial t$ of Eq. (8) into Eq. (6), there will be two independent trans-sharp-interface fluxes [30], whose changes in free energy are

$$\Delta f_d^I = h_\alpha\left(\tilde{\mu}_\alpha - \tilde{\mu}_\gamma\right)\left(c_{trans}^{SI} - c_\gamma\right), \tag{11}$$

$$\Delta f_d^{II} = h_\gamma\left(\tilde{\mu}_\alpha - \tilde{\mu}_\gamma\right)\left(c_{trans}^{SI} - c_\alpha\right), \tag{12}$$

and $\Delta f_m + \Delta f_d^I + \Delta f_d^{II}$ is the same as Eq. (9) from three conjugated forces and fluxes. Additionally, Eqs. (11) and (12) can be further extended as Part I: $\Delta f_d^I\big|_{1.2} = h_\alpha\tilde{\mu}_\alpha\left(c_\alpha - c_\gamma\right)$, $\Delta f_d^I\big|_{1.3} = h_\alpha\tilde{\mu}_\alpha\left(c_{trans}^{SI} - c_\alpha\right)$, $\Delta f_d^I\big|_{1.4} = h_\alpha \tilde{\mu}_\gamma (c_\gamma - c_{trans}^{SI})$ ; and Part II: $\Delta f_d^{II}\big|_{2.1} = h_\gamma \tilde{\mu}_\gamma (c_\alpha - c_\gamma)$ , $\Delta f_d^{II}\big|_{2.2} = h_\gamma \tilde{\mu}_\gamma (c_\gamma - c_{trans}^{SI})$ , and $\Delta f_d^{II}\big|_{2.3} = h_\gamma \tilde{\mu}_\alpha (c_{trans}^{SI} - c_\alpha)$, whose physical meaning will be illustrated later.

Now, we interpret the Eqs. (10)-(12) with the following "two-part" phase transformation mechanisms, whose schematic depiction is shown in Figure. 2(a), and Figure. 2(b, c) exhibit the free energy changes of different dissipative processes.

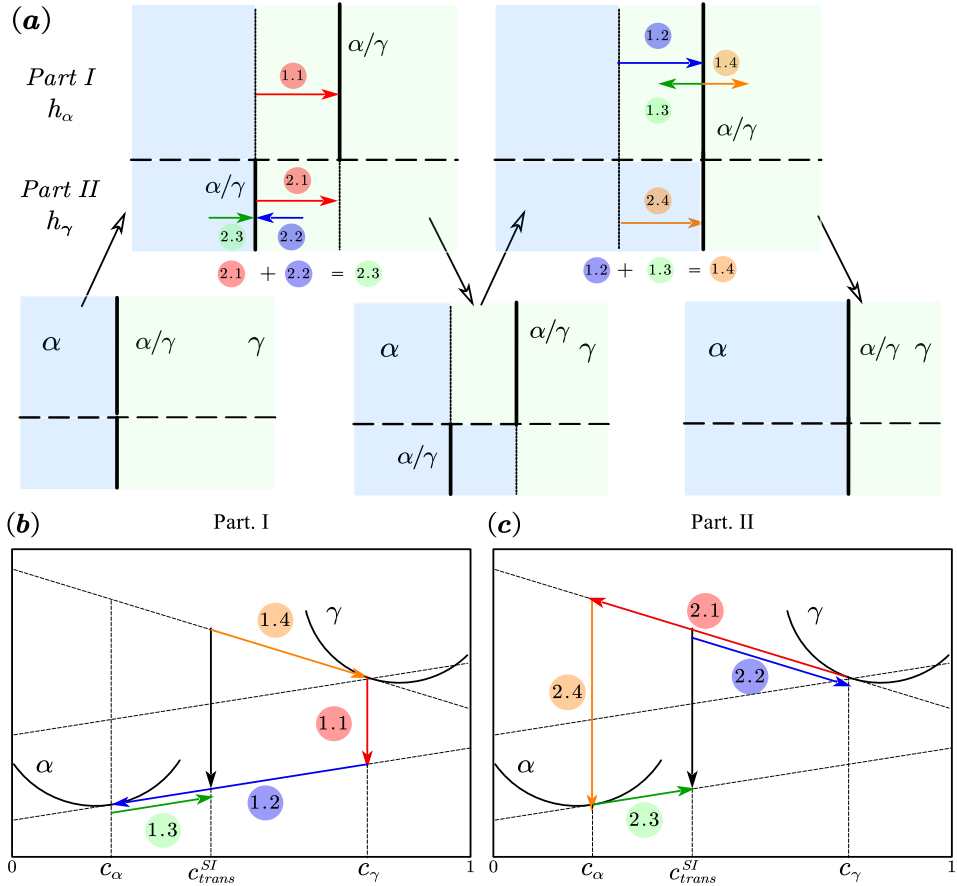

FIG. 2. (a): The "two-part" schematics for "sharp-interface" type phase transformation in RVEs; (b) and (c): Graphical constructions of molar free energy changes for Parts I and II, respectively. Parts I and II are divided by the dashed lines in (a). In each part, the phase boundary is represented by the black solid line, while the background colors represent the lattice concentrations of $\alpha$ and $\gamma$. The (1.1-1.4) and (2.1-2.4) denote the kinetic processes of Part I and II, respectively.

For Part I ($h_\alpha$), the phase boundary first diffusionless transform from $\gamma(c_\gamma)$ to $\alpha(c_\gamma)$ [(1.1) in Figure. 2], during which two trans-sharp-interface fluxes and their corresponding free energy changes are compensated, according to Eq. (6). Then, the metastable $\alpha(c_\gamma)$ starts desorbing atoms to approach $\alpha(c_\alpha)$ [(1.2)], which also produces a moving tendency of the phase boundary. Concurrently, the small changes in concentration and phase boundary induce two non-compensated trans-sharp-interface fluxes [(1.3) and (1.4)], providing an opposite moving tendency of the phase boundary. As a result of (1.2)-(1.4), the phase boundary is not moved. Alternatively, the net material transferred from product $\alpha$ to the sharp interface equals that transferred from the interface to parent $\gamma$, i.e., $(1.2)+(1.3)=(1.4)=c_\gamma - c_{trans}^{SI}$, so there are no retained atoms moving the phase boundary. The net free energy change, Eq. (11), is actually the same as Hillert et al.'s work [12].

For Part II ($h_\alpha$), similarly, (2.1)-(2.3) first make $\gamma(c_\gamma)$ desorb atoms into $\gamma(c_\alpha)$, while the phase boundary is static due to the offset by $(2.1)+(2.2)=(2.3)=c_\alpha - c_{trans}^{SI}$. The net change in free energy is given by Eq. (12). Then, the phase boundary migration makes the metastable $\gamma(c_\alpha)$ phase transform into $\alpha(c_\alpha)$ in a diffusionless way [(2.4) in Figure. 2]. Note that previous sharp-interface models [11-15] can also be reinterpreted by the present "two-part" mechanism, whose proportion is determined by atomic mobility for trans-sharp-interface diffusion and long-range diffusion fluxes at the phase boundary [30]. Additionally, a recent experiment by neutron total scattering [5] should be mentioned. It is found that there exist "atomic groups or cells" in the interface, analogous to the present RVE concept, and the phase transformation shows a diffusion-assisted displacive manner as the Part II.

*Thermodynamics of diffuse-interface*.—Now, it is necessary to explore the relationship between the present PF model with the classical diffuse interface theories by Cahn [31] and Hillert and Sundman [32]. Using a standard coordinate transformation $\partial/\partial t = V \partial/\partial x$ for the steady state, Eq. (1) is transformed into $J_{steady} = -(V/v_m)(\hat{c} - \hat{c}_{trans}^{DI})$. $\hat{c}_{trans}^{DI}$ is the conserved concentration at a reference point, at which $J = 0$. Then, apply $\partial/\partial t = V \partial/\partial x$ to Eqs. (5)-(7) and multiply by $\partial\phi/\partial x$, $\partial c_{i=\alpha,\gamma}^*/\partial x$, and $\hat{J}_{steady}$ on both sides. After integrating the whole diffuse interface, the balance of the (absolute) total driving force $\Delta f_{tot}^{DI}$ and energy dissipation $Q_{tot}$ is given by

$$\Delta f_{tot}^{DI} = f_\gamma^\delta - f_\alpha^{-\delta} + \tilde{\mu}_\gamma^\delta \left( \hat{c}_{trans}^{DI} - c_\gamma^\delta \right) - \tilde{\mu}_\alpha^\delta \left( \hat{c}_{trans}^{DI} - c_\alpha^{-\delta} \right)$$
$$= \int_{-\delta}^{+\delta} \left\{ \frac{V}{M_\phi} \left( \frac{\partial \phi}{\partial x} \right)^2 + \sum_{i=\alpha,\gamma} \frac{V}{M_i^*} \left( \frac{\partial c_i^*}{\partial x} \right)^2 + \frac{1}{V} \frac{J^2}{M_c} \right\} dx = Q_{tot}. \quad (13)$$

The superscripts "$-\delta$" and "$\delta$" denote the $\alpha$ and $\gamma$ boundaries of the diffuse interface, respectively. It is shown that $\Delta f_{tot}^{DI}$ is analogous to that of sharp-interface $\Delta f_{tot}^{SI}$, and $\hat{c}_{trans}^{DI}$ can be similarly called as the material composition crossing the diffuse interface. If assuming the effective phase-field mobility $M_\phi^{eff}$ in Eq. (5) as a constant, Eq. (13) can be further simplified by combining Eq. (2), yielding as

$$Q_{tot} \approx \frac{V}{\left( \kappa^2 / \sigma_{\alpha\gamma} \right) M_\phi^{eff}} + \int_{-\delta}^{+\delta} \left\{ \frac{1}{V} \frac{J^2}{M_c} \right\} dx, \quad (14)$$

where $\sigma_{\alpha\gamma}$ is the interface energy. One can see that the first and second terms on the right-hand side respectively correspond to the interface friction/migration ($Q_m$) and solute drag ($Q_d$) of the classical diffuse interface theories [31-33]. Moreover, $M_\phi^{eff}$ is related to the intrinsic interface mobility $m$ by $M_\phi^{eff} = m \left( \sigma_{\alpha\gamma} / \kappa^2 \right)$.

*Partial-drag in rapid solidification.* To validate the present PF model in representing non-equilibrium interface kinetics, we apply it to the 1-D rapidly solidified Al-Cu alloys. Almost all model parameters [30] are the same as Haapalehto et al.'s molecular-dynamic (MD) simulations [21], e.g., phase diagram, diffusion coefficients, interface thickness, and partition coefficient $k(V) = c^{max} / c^{-\delta}$. Figure. 3(a) and (b) exhibit the concentration profiles and their corresponding non-equilibrium $k(V)$, where there is an increasing solute trapping with interface velocity $V$. The complete solute trapping at a finite velocity $V_c$ is reproduced by modifying the $M_c$ of Eq. (7) with an effective coefficient $1 + \left( v_m / V_c^2 \right) \left( 1 / \nabla c \right) \left( \partial J / \partial t \right)$ [30], which accounts for the far-from-equilibrium diffusion in the very high velocity regime [34]. After reaching $V_c$, the diffusionless solidification occurs, where the solute drag $Q_d$ disappears in Figure 3(c), and thus $V$ is linearly proportionate to the undercooling in Figure 3(d).

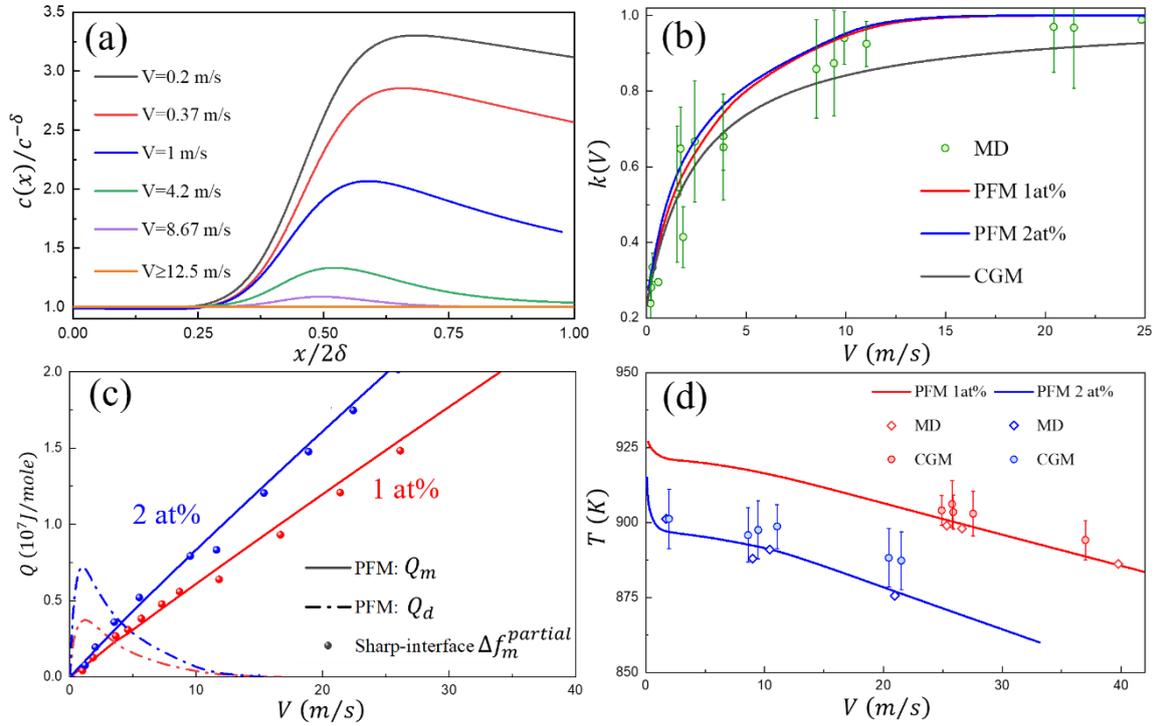

FIG. 3. (a) and (b): Present composition profiles and corresponding partition coefficient; (c) and (d) Energy dissipation and interface temperature. Haapalehto et al.'s results [21] of MD simulations and sharp-interface model (CGM) are listed for comparison.

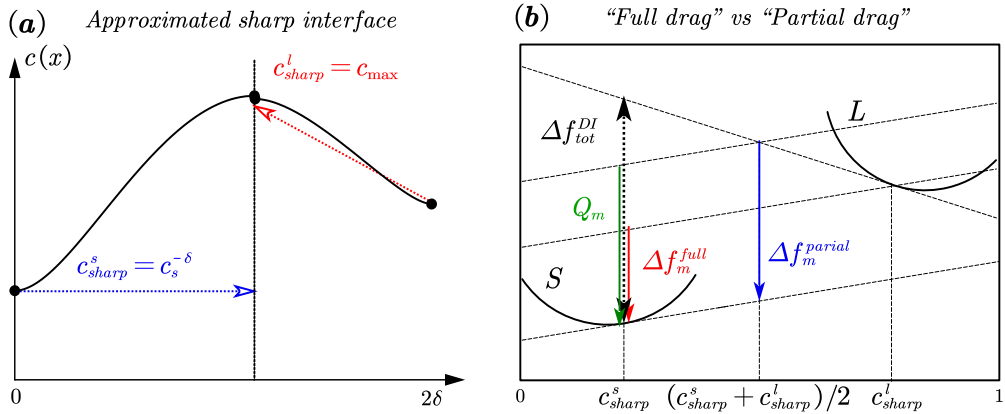

FIG. 4. (a) Schematic depiction of an approximated sharp interface for present PF model; (b) Comparison of simulated $Q_m$ with "full drag" and "partial drag" models [19].

Finally, we can revisit the atomic simulated "partial drag" phenomenon [19-22] with the present diffuse interface model. We starts from a "sharp-interface" approximation with $c_l^{sharp} = c^{max}$ and $c_l^{sharp} = c^{-\delta}$ in Figure. 4(a). Because of the negligible diffusion in bulk solids, $\hat{c}_{trans}^{DI} = c^{-\delta}$ in Eq. (13). The corresponding $\Delta f_{tot}^{DI}$ is thus given by the black arrow in Figure 4(b) as in the previous full-drag sharp-interface models [9-14]. However, as shown in Figures. 3(c) and 4(b), the simulated $Q_m$ is very close to "partial-drag" sharp interface model [9], i.e., $\Delta f_m^{partial} = f_s - f_l - 0.5(\tilde{\mu}_s + \tilde{\mu}_l)(c_s - c_l)$. The coefficient "0.5" surprisingly coincides with the value recently summarized from almost all existing atomic simulations [22]. It indicates that the present work can reproduce the non-equilibrium interface conditions self-consistently in thermodynamics, while the recent "partial drag" sharp interface model [11] is a convenient method for approximation. In the future, it is important to explore a strategy of quantitative extending the present PF model to a size scale available for mesoscale simulations.

The work was supported by the Research Fund of the State Key Laboratory of Solidification Processing (NPU), China (Grant No. 2020-TS-06, 2021-TS-02) and Natural Science Basic Research of Shannxi (Program No. 2022JC-28).